\documentclass[12pt,preprint]{aastex}

\shorttitle{C$_5$N$^-$ in IRC +10216}

\shortauthors{Cernicharo et al.}

\begin{document}

\title{Detection of C$_5$N$^-$ and vibrationally excited C$_6$H in
IRC +10216 \thanks{
Based on
observations carried out with the IRAM 30-meter telescope. IRAM is
supported by INSU/CNRS
(France), MPG (Germany) and IGN (Spain).}}

\author{J. Cernicharo\altaffilmark{1}, M. Gu\'elin\altaffilmark{2},
M. Ag\'undez\altaffilmark{1}, M. C. McCarthy\altaffilmark{3},
and P. Thaddeus\altaffilmark{3}}

\altaffiltext{1}{DAMIR, Instituto de Estructura de la Materia, CSIC, Serrano 121, 28006
Madrid, Spain; cerni@damir.iem.csic.es}
\altaffiltext{2}{Institut de Radioastronomie Millim\'etrique, 300 rue de la
Piscine, 38406 St. Martin d'H\'eres,France}
\altaffiltext{3}{Harvard-Smithsonian Center for Astrophysics, 60 Garden
Street, Cambridge, MA 02138, U.S.A. and School of Engineering and
Applied Sciences, Harvard University, 29 Oxford Street, Cambridge,
MA 02138, U.S.A.}

\vspace{3.0cm}
Accepted for publication in the Astrophysical Journal Letters Oct/08/2008.

\begin{abstract}
We report the detection in the envelope of the C-rich star IRC+10216
of four series of
lines with harmonically related frequencies: B1389, B1390, B1394 and
B1401. The four series
must arise from linear molecules with mass and size close to those of
C$_6$H and
C$_5$N. Three of the series have half-integer rotational quantum
numbers;
we assign them to the
$^2\Delta$ and $^2\Sigma^-$ vibronic states of C$_6$H in its lowest
($\nu_{11}$) bending
mode. The fourth series, B1389, has integer J with no evidence of
fine or hyperfine structure; it   has a rotational constant of
1388.860(2)~MHz and a centrifugal distortion
constant of 33(1)~Hz; it is almost certainly the C$_5$N$^-$ anion.

\end{abstract}


\section{Introduction}

The dusty envelope  of the  carbon star IRC+10216 is a rich radio 
source where at least 70
molecules have been observed (see, e.g., Cernicharo et al. 2000).
Particularly rich in long
linear carbon chains, it is the source where the polyacetylenic radicals C$_n$H
($n=4,6,8$; see Gu\'elin et al., 1987;
Cernicharo \& Gu\'elin 1996) and C$_n$N (n=3,5; see
Gu\'elin et al., 1998) were first detected and identified. The
butadiynyl
radical C$_4$H has been observed there in rotational lines of the
ground and several
vibrationally-excited states. The
excited state lines are strong, owing
to perturbation by the low-lying $A^2\Pi$
electronic state (Yamamoto et al. 1987).
Notably poor in cations,
IRC+10216 
has turned out to be
   the richest source of anions in the sky. All known interstellar anions,
C$_6$H$^-$, C$_4$H$^-$, C$_8$H$^-$, and
C$_3$N$^-$ (Thaddeus et al. 2008 and references therein), have been 
identified there,  where they
are fairly abundant. The presence of
carbon chain negative ions in space was predicted long ago (see e.g. 
Sarre 1980, Herbst
1981)
on the ground that electron radiative attachment is
efficient for molecules with large electron
affinities and dense vibrational spectra. The presence of C$_3$N and
C$_5$N in IRC+10216 and the recent detection of
C$_3$N$^-$ in this source, imply that C$_5$N$^-$ may be present.

\section{Observations}
The astronomical observations presented here are mainly from a
spectral line survey of IRC+10216
done with the IRAM 30-m telescope between 1995 and 2008 (Cernicharo
et al. 2000;
2008 in preparation). The 3-mm part of this survey covers the 80-115 
GHz band with an
r.m.s. noise of 0.3-2 mK per 1 MHz-wide channel.
Data were taken with the secondary nutating by
$90''-120''$, an arrangement which produced
very flat spectral baselines. Two SIS 3-mm
receivers, with orthogonal polarizations and system temperatures of
$100-130\;$K, were used
simultaneously. Pointing and focus were regularly checked on planets
and on the strong
nearby quasar OJ~287.
In Fig. 1 \& 2 the intensities are given in
$T_A*$, the antenna temperature corrected for atmospheric absorption 
and rear spillover
losses, by means of the ATM code (Cernicharo 1985). A few 
complementary observations at 22 GHz were carried out in March 2008 
with the
MPIfR Effelsberg telescope, both on IRC+10216 and on TMC--1.
Derived line parameters are given in Table 1.

\section{Results: Vibrationally excited C$_6$H}
\begin{figure}
\includegraphics[angle=0,scale=1.0]{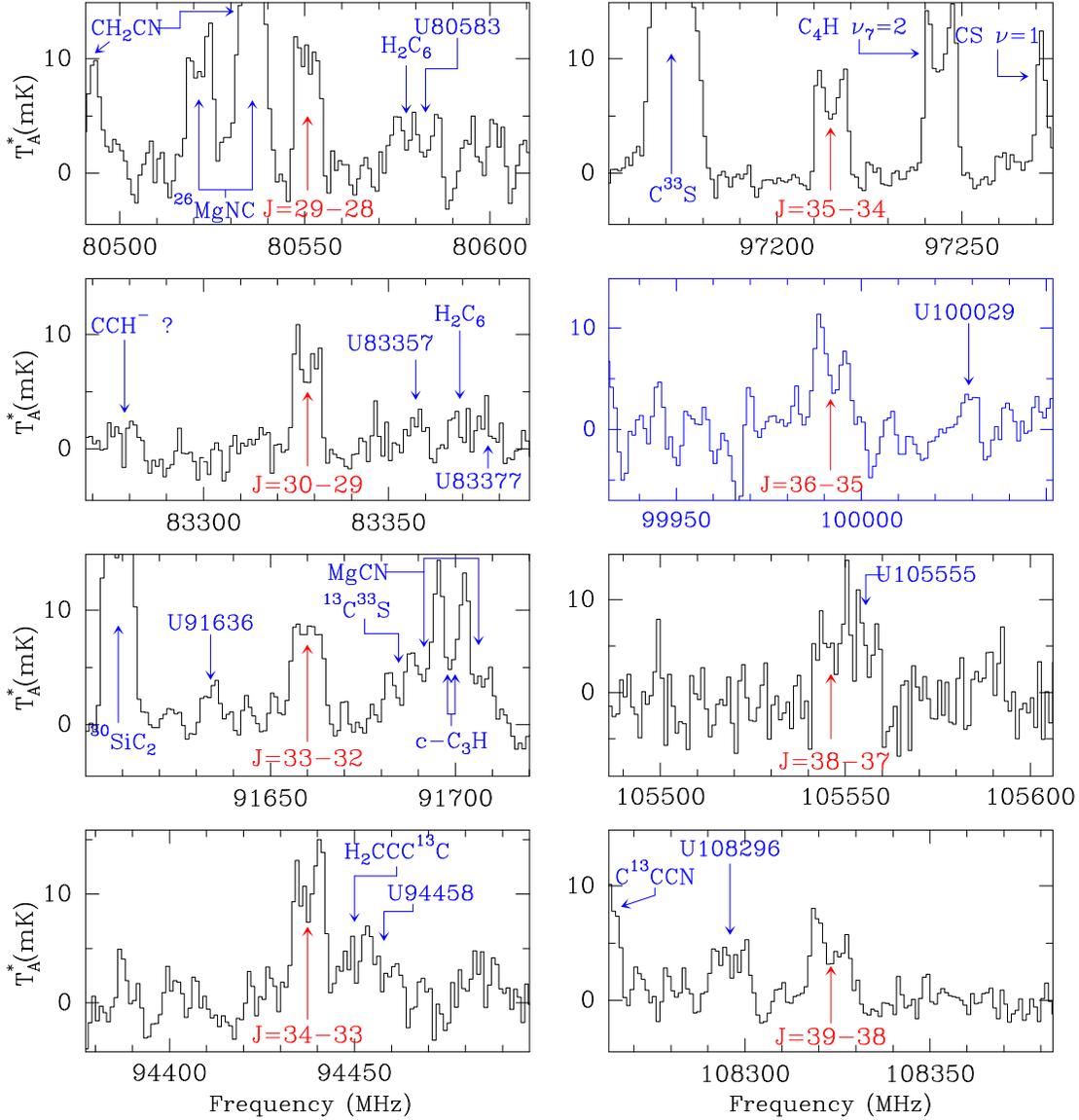}
\caption{Spectra of IRC +10216 observed with the IRAM 30-m
telescope, showing lines from the B1389 series
assigned here to C$_5$N$^-$. The marginal weak line U83278 is worth 
noting, because it is within 0.1
MHz of the J=1-0 line of CCH$^-$ (see text).}
\label{B1389} \vspace{-0.1cm}
\end{figure}

\begin{figure}
\includegraphics[angle=-90,scale=1.0]{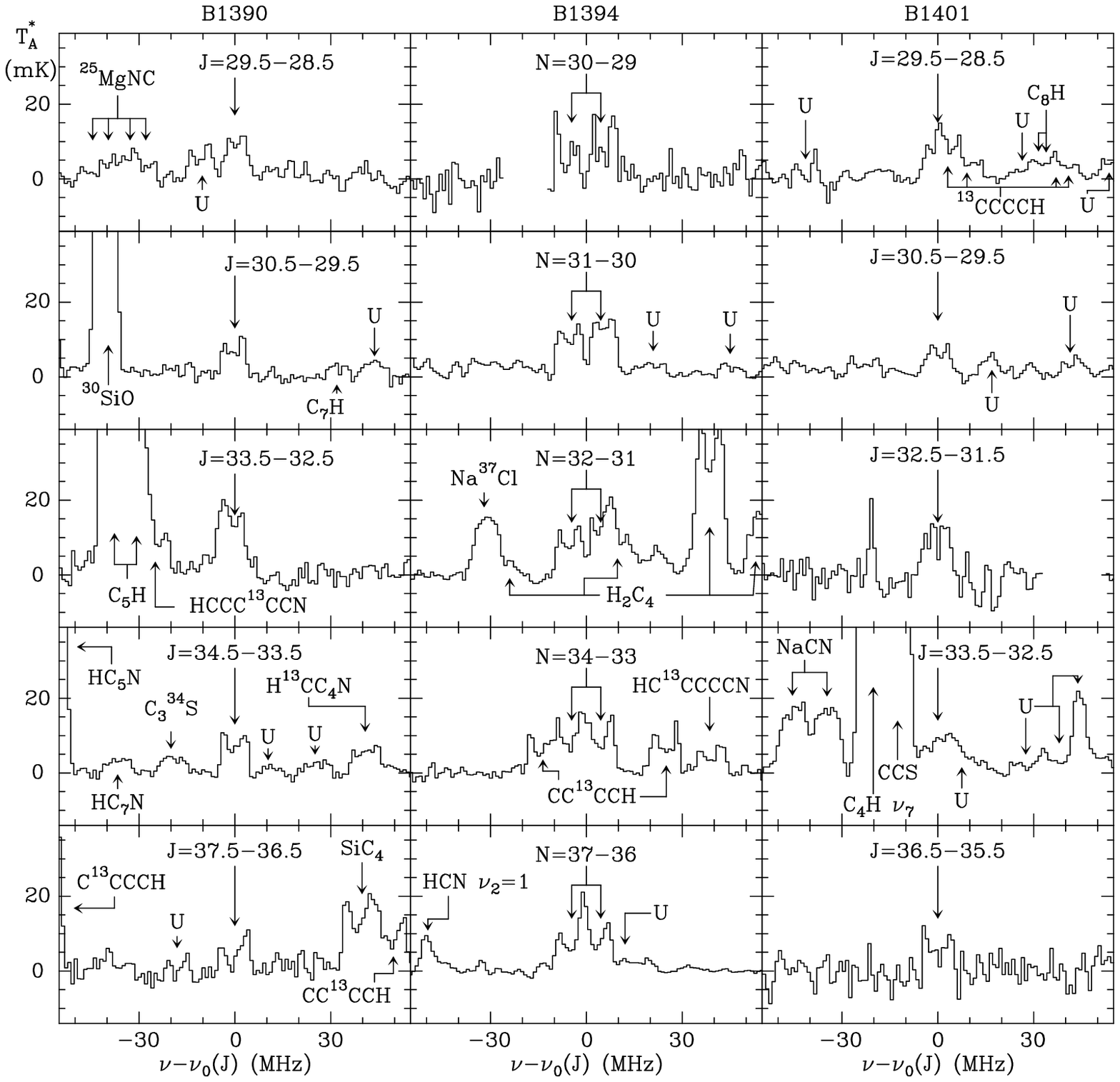}
\caption{Spectra of IRC +10216 observed with the IRAM 30-m
telescope showing selected lines pertaining to the series B1390, B1394
and B1401. These three series of lines are
assigned to vibronic states of the $\nu_{11}$ bending
mode of C$_6$H.}
\label{BBs} \vspace{-0.1cm}
\end{figure}

During the survey, about 1500 lines in the 80-115 GHz
band were observed, most from known molecules. A
few hundred escaped identification; among these, the four harmonic
series with either integer or half-integer quantum numbers in 
Fig~\ref{B1389} and Fig~\ref{BBs} stand out; for one series, B1394, 
the lines are split into doublets (see Tables 1 \& 2).
Since the B and D values are very close to those of C$_6$H and C$_5$N
(B= 1391.2 MHz, D= 51 Hz, and B= 1403.1 MHz, D= 50 Hz, respectively),
the carriers of
the four series must be linear molecules of weight and
size similar to that of these two radicals.

C$_6$H is a linear radical with a $^2\Pi$ ground electronic state
(Cernicharo et al., 1987). It has 6 stretching modes, $\nu_1-\nu_6$ and
5 bending modes,
$\nu_7-\nu_{11}$. The modes $\nu_{10}$ and $\nu_{11}$ have the lowest
energies, $\simeq$210
and $\simeq$93 cm$^{-1}$, respectively, according to {\it
ab-initio} calculations by Cao and Peyerimhoff (2001);
the next mode $\nu_9$ is at about 400
cm$^{-1}$.

Even though excited vibrational states of long carbon chains are
susceptible to infrared fluorescent excitation in IRC+10216,
vibrational temperatures remain low: e.g., $40-50$ K for
the $\nu_7$ mode of HC$_3$N and the $\nu_{11}$ mode of HC$_5$N
(Cernicharo et al. 1987, 2008).  Only the lowest
bending mode of C$_6$H ($\nu_{11}$) is therefore expected to be sufficiently
populated to detect.
The first excited electronic state of C$_6$H, $^2\Sigma$, is known to
lie very close in energy to the $^2\Pi$ ground state (Murakami et al., 1987).
Owing to coupling between the
vibrational bending
modes and the electronic orbital motion of the molecule (the Renner-Teller
effect), each bending mode of C$_6$H is split into one
$^2\Delta$ (with
half-integer quantum numbers) and two $^2\Sigma$ vibronic states
($^2\Sigma^+$ and
$^2\Sigma^-$ with integer quantum numbers).
According to Herzberg (1966), the
splitting between the $\Sigma^+$ and $\Sigma^-$ states, A$_{\Sigma}$,
produced by the
Renner-Teller effect is \\
(1) A$_{\Sigma}$=(A$_{SO}^2$+4$\epsilon^2\omega^2$)$^{1/2}$,\\
where A$_{SO}$ is the spin-orbit
constant of the ground $^2\Pi$ electronic state of C$_6$H
(A$_{SO}$=-15.04 cm$^{-1}$, Linnartz et al., 1999), $\epsilon$ is the
Renner parameter, and $\omega$
is the frequency of the bending mode. The $^2\Sigma^-$ and $^2\Sigma^+$
states lie above and below the $^2\Delta$ state by A$_{\Sigma}$/2.
The values of $B$ and $\gamma$ for the $^2\Sigma^\pm$
states, the
effective spin-orbit constant for the $^2\Delta$ vibronic state,
and the effective
rotational constants for the two ladders of this state are given by
(Herzberg, 1966),\\
(2) $B^\pm_\Sigma$=B(1 $\pm$ A$_{SO}^2$B/A$_\Sigma^3$), \\
(3) $\gamma^\pm_\Sigma$=2B(1-2$\epsilon\omega$/A$_\Sigma$ $\pm$
A$_{SO}^2$B/A$_\Sigma^3$), \\
(4) A$_{eff}$=A$_{SO}$(1-3$\epsilon^2$/4),\\
(5) B$_{eff\Delta}$=B(1$\pm$B/A$_{eff}$),\\
where B is the unperturbed rotational constant of the bending mode. 
These expressions have to be taken with caution as additional
ro-vibronic coupling with higher lying vibronic states may introduce 
significant
higher-order corrections (see, e.g., Brown, 2003).

In view of the close agreement between the rotational constant of
C$_6$H and those of our four new series, it 
was probable                  
that one or 
more may by produced by the low-lying
$\nu_{11}$ bend of C$_6$H. By analogy with HC$_5$N, which
has a
vibrational spectrum similar to that here, C$_6$H in its $\nu_{11}$ 
bending mode is
expected to have a rotational
constant 1.004 times larger than that in its vibrational ground 
state, i.e. close to 1395 MHz.
The assignment of
B1390 and B1401, both with half-integer quantum numbers and whose average
B value is close
to 1395 MHz, to the $^2\Delta_{3/2}$ and
$^2\Delta_{5/2}$ ladders of the
$^2\Delta$ vibronic state of $\nu_{11}$ was therefore 
logical. As 
discussed below, these assignments have
been confirmed in the laboratory (see below).

Because B1394 shows well resolved doublets with integer rotational quantum
numbers (see Fig~\ref{BBs}), this series is almost certainly from a 
$^2\Sigma$ vibronic state. The most
likely carrier is the lowest of
the two $^2\Sigma$ ladders of the $\nu_{11}$ state,  the $^2\Sigma^-$
ladder. We note that
we have found no other series of line doublets of comparable or
greater intensity that could
match this ladder, would B1394 pertain to another
vibrational state.

With the assignment of B1394 and B1390/B1401 series to the
$^2\Sigma^-$ and the
$^2\Delta$ ladders of $\nu_{11}$, we can derive from relations (1)
through (5) a set of
ro-vibrational parameters that give a good fit to the data:
B$_\Delta$=1395.72
MHz, A$_{eff_\Delta}$=11.13 cm$^{-1}$,
$\epsilon^2$=0.352, A$_\Sigma\simeq$145.0
cm$^{-1}$ and $\omega\simeq$120 cm$^{-1}$, a value compatible with
that calculated for $\nu_{11}$
by Cao and Peyerimhoff (2001).
Therefore, under the assumption that relations (1) through (5) apply
to C$_6$H,
the $^2\Sigma^-$, $^2\Delta$, and $^2\Sigma^+$
vibronic states lie
68, 173 and 277 K above the ground state.
The $^2\Sigma^+$ vibronic state will
have a rotation
constant (B$^+$) so close to that of the $^2\Sigma^-$ state (B$^-$),
and a doublet
separation so similar to $^2\Sigma^+$ ($\gamma^+\simeq\gamma^-$) that
its rotational lines
will be blended in IRC+10216 with those of $^2\Sigma^-$ up to very
high J  (where lines are probably too weak to detect).
The column densities derived for the
$\nu_{11}$ vibronic
states are: N($^2\Sigma^-$)$\simeq$1.4$\times$N($^2\Delta$)=8
10$^{12}$ cm$^{-2}$. Both
vibronic states have a rotational temperature of $\simeq$45 K ---
somewhat higher than that
measured for the ground vibrational state (31$\pm$2 K; Cernicharo
et al., 2007). The column density in
the ground state is calculated to be $6.6\times 10^{13}$ cm$^{-2}$, 
indicating that about 20\% of
C$_6$H is in
the $\nu_{11}$ state.

Recent laboratory measurements at both millimeter- and
centimeter-wavelengths confirm that B1390, B1401, and B1394 arise from
$^2\Delta$ and $^2\Sigma$ components of a low-lying vibrationally
excited state of C$_6$H (Gottlieb et al. 2006).  The $^2\Sigma$
component has been observed over a wide range of rotational
excitation ($N$=5 to 85), under experimental conditions which optimize
lines of ground state C$_6$H.  To reproduce the observed spectrum, an
effective Hamiltonian with several higher-order terms in the
spin-rotation constant is required.  Although weaker by a factor of
three or more relative to B1394, the high-lying $^2\Delta$ component
has also been observed in the same laboratory discharge.  The
spectroscopic constants derived from the astronomical data agree to
within $1\sigma$ with those derived from the larger and more precise
set of laboratory data for both the $^2\Delta$ and $^2\Sigma$
components. A complete account of the laboratory observations will be
given elsewhere.

There is so far no laboratory evidence for the remaining
astronomical series B1389 under conditions where lines of ground
state and vibrationally excited C$_6$H or ground state C$_6$H$^-$ are
observed.

\section{Assignment of B1389 with C$_5$N$^-$}

The remaining series, B1389, has integer J and
exhibits no evidence of hyperfine or other structure (see the 
J=$35-34$ line in Fig. 1). For this reason, and because of the 
strength of these lines, it cannot be assigned to any of the bending 
modes of C$_6$H. Its carrier is almost certainly a new linear 
molecule with a $^1\Sigma$
ground electronic state, and a molecular weight close to that of 
C$_6$H and C$_5$N.

The most obvious candidates are the ions C$_6$H$^-$, C$_6$H$^+$ and
C$_5$N$^-$.
C$_6$H$^-$ has already been detected in the laboratory and in 
IRC+10216 (McCarthy et al., 2006), and is immediately ruled out.
Linear C$_6$H$^+$ has a $^3\Sigma$ ground electronic state (according
to Fehr\'er \& Maier, 1994), and only a fraction of its rotational 
transitions will be harmonically related with B$\simeq$1340 MHz.
Moreover, only one cation, HCO$^+$,
has so far been detected in IRC+10216, and it has a very low
abundance.
We therefore conclude that  C$_6$H$^+$ is not the carrier of B1389.

Finally, C$_5$N$^-$ has the right
$^1\Sigma$ ground electronic state and calculated rotational and
distortion constants: 1389 MHz and 33 Hz (Botschwina, 2008,
private communication; see also
Aoki 2000). Botschwina quotes an error as small
as .5\% for the
rotational constant. The match with B1389 is perfect, and it is therefore
tempting to conclude that C$_5$N$^-$ is our new molecule.

\begin{figure}
\includegraphics[angle=-90,scale=.68]{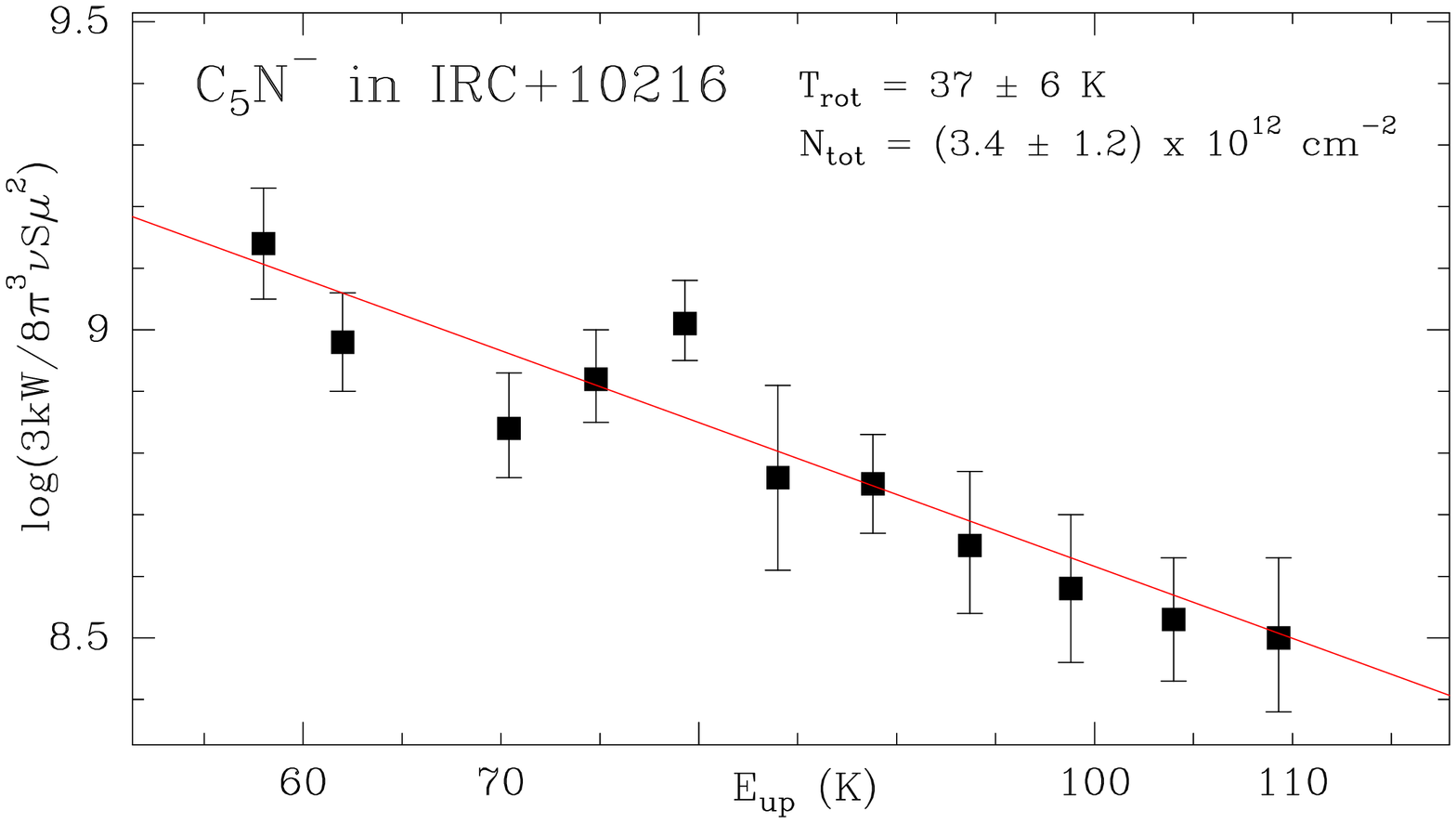}
\caption{Rotational diagram for B1389. The column density
has been derived assuming a dipole moment of 5.2 D (see text).}
\label{C5N-} \vspace{-0.1cm}
\end{figure}

The only reservation we might see 
to this identification is the
intensity of the lines, which in IRC+10216 are about 
twice 
stronger than those of
the parent species
C$_5$N. Supporting the identification, the lines of the anion benefit 
from a more
favorable partition function
(no doublets) and from a larger permanent dipole moment: the latter
has been calculated by
Botschwina (2008) to be 5.2 D. Assuming our identification is
correct, we derive a rotation
temperature of 37$\pm$6 K and a C$_5$N$^-$ column density of
3.4$\times$10$^{12}$ cm$^{-2}$
(see Fig~\ref{C5N-}). Neutral C$_5$N has been observed in IRC+10216
by Gu\'elin et al.
(1998), who derived a column density of 6$\times$10$^{12}$ cm$^{-2}$
for a dipole moment of
3.39 D (Botschwina 1996).
The abundance ratio between the neutral
and the anion is then only 1.8, the largest relative abundance
observed so far for any
  anion.

A number of clues, however, suggest that we may have underestimated 
the 
abundance of
neutral C$_5$N. First of all, as noted by Gu\'elin et al.~(1998), the abundance
decrement C$_3$N/C$_5$N is an order of magnitude larger than that
of HC$_3$N/HC$_5$N and
C$_4$H/C$_6$H, both in IRC+10216 and in TMC-1 --- yet all chemical
models predict it to be
similar. Second, in the laboratory the lines of C$_5$N are
far weaker than those
of C$_6$H, yet the production of both should be similar.
Pierre Valiron (private
communication 2008) has re-calculated the ground state symmetry and 
dipole moment of C$_5$N at
restricted open-shell Hartree-Fock
and coupled cluster level,  using augmented correlation consistent
Dunning's basis sets. Using a double zeta set, he derives a $^2\Pi$ 
ground state with
a small ($\sim 1$ D) dipole moment, in agreement with the previous
calculations of Pauzat et al.~(1991); alternatively with
a quadruple zeta set, he finds a $^2\Sigma$ ground state and a
large dipole moment ($\simeq$ 3.4 D), in agreement with Botschwina
(1996). Valiron concludes that the dipole
moment of C$_5$N may well lie between those two values in the case of
admixing between the $^2\Sigma$ and $^2\Pi$ states similar to those
observed for C$_4$H and C$_6$H. A dipole moment twice smaller than
that calculated for the unperturbed $^2\Sigma$ state would raise the
C$_5$N/C$_5$N$^-$  abundance ratio to 8, making it only slightly
smaller than
the C$_6$H/C$_6$H$^-$ ratio
(see Cernicharo et al. 2007).

\begin{figure}
\includegraphics[angle=-90,scale=.62]{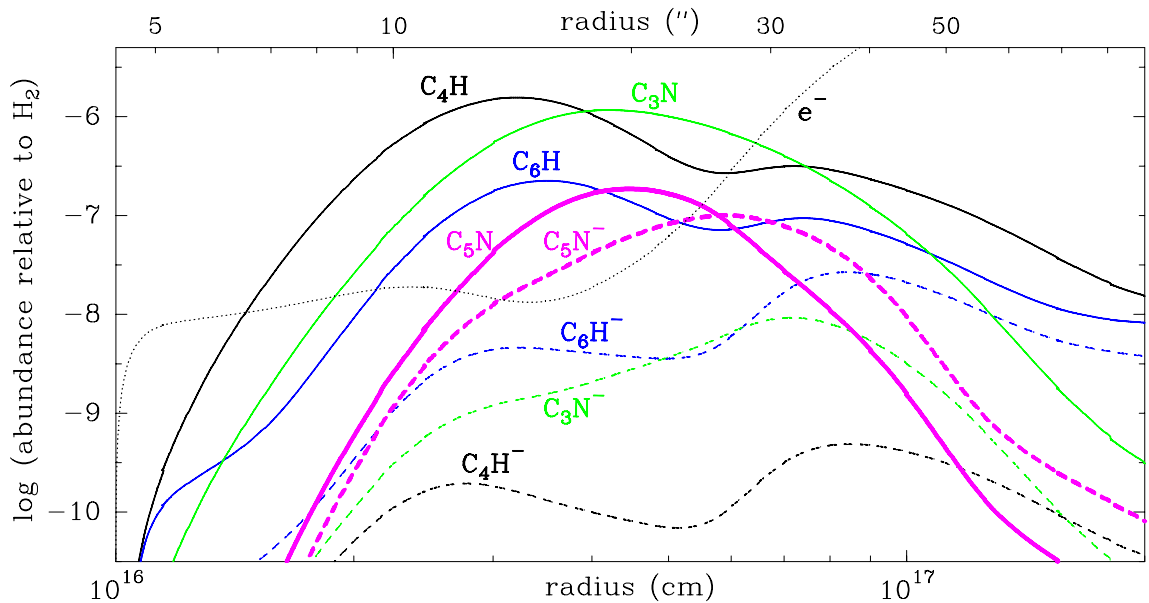}
\caption{Model abundances of the neutral radicals
C$_n$H, C$_n$N and their anions in the outer envelope of
IRC+10216.}
\label{AC5N-} \vspace{-0.1cm}
\end{figure}

To further investigate the formation of C$_5$N
and C$_5$N$^-$, we have modeled the chemistry in the
external layers of the circumstellar envelope of IRC+10216 with the
same time-dependent
chemical model we have used to calculate the abundances of C$_4$H and
C$_4$H$^-$ (Cernicharo
et al., 2007) 
and C$_3$N and C$_3$N$^-$(Thaddeus et al., 2008).
Fig~\ref{AC5N-} shows the
computed abundance as functions of the distance to the star, $R$. In
these models we have
assumed an electron radiative attachment rate for C$_5$N of
2$\times$10$^{-7}\times$(T/300)$^{-0.5}$ cm$^3$ s$^{-1}$, 
close to those 
adopted for the largest C$_n$H radicals.
In the region of interest ($R= 3-5 \times
10^{16}$ cm), C$_5$N is found to be 2-10 times more abundant than
C$_5$N$^-$, which is in
good agreement with our identification of B1389 with C$_5$N$^-$.
We note that our spectral survey should have been sensitive enough
to detect another series with a rotational constant close to that predicted
by Botschwina, assuming that B1389 does not arise from this anion and 
that the dipole moment
of C$_5$N$^-$ is as large as 5.2 D. No such series was found.

Laboratory searches to detect C$_5$N$^-$ have also been
undertaken with the microwave spectrometer at Harvard on the assumption that
this anion is the carrier of B1389. No frequency search is required,
because the rotational constant derived from the astronomical data is
good to a few kHz, and the nitrogen hfs is negligible throughout much
of the centimeter-wave band. As for C$_6$H$^-$, experimental
parameters were first optimized to produce strong lines of the
corresponding neutral radical, C$_5$N.  A number of searches for
C$_5$N$^-$ were then performed using different discharge voltages
(ranging from 1000 V to 600 V); still other searches using a range of
concentrations and gas pulses of various lengths, were also attempted,
all without success.
Our failure to detect C$_5$N$^-$ is not particularly disturbing,
because finding C$_4$H$^-$ and C$_3$N$^-$ was very difficult
(McCarthy \& Thaddeus 2008), requiring repeated trial and error. 
Additional experiments are needed to find the highly specific
conditions which are apparently required to make this anion.

Finally, a number of other candidates for B1389 were considered and 
rejected, including
AlCCCN, AlNCCC, NaCCCN, and
NaNCCC.  All have $^1\Sigma$ electronic states, large dipole moments
($\mu_a$=6.2, 4.0,
13.6, \& 16.6 D, repectively), and about the right mass and size.
The calculated rotational
constants, however, differ by
more than 4\% from 1389 MHz, versus an estimated uncertainty of 1\%
(Petrie 1999) --- probably a prohibitive discrepancy.
Finally,
MgCCCN, whose calculated rotational constant  is close to B1389 (1397
MHz -- see
Petrie 1999), has a $^2\Sigma$ not a $^1\Sigma$ ground electronic
state.

With no plausible alternative, we therefore conclude that
B1389 is almost certainly C$_5$N$^-$.  The observation of low-J lines in the
cold source TMC-1  
would strengthen this conclusion. We
have searched for the
J=8-7 transition in TMC-1 with the Effelsberg telescope, and have
detected a 3$\sigma$
feature at the right frequency; a further factor of two
improvement in sensitivity is required for confirmation. The final
proof, however, must await
detection of the B1389 lines in the laboratory.

Among the unidentified lines of Fig. 1, we note that U83278 is within
0.1 MHz of the J=1-0 line of
CCH$^-$. Assuming the line arises from this anion,
we calculate an abundance ratio CCH/CCH$^-$ $\simeq$12500.
The J=3-2 CCH$^-$ transition  would
then be strong enough to be detectable; a search for this transition to confirm
the present assignment is under way.

\vspace{-0.53cm}
\acknowledgments

Shortly before this Letter was submitted, Pierre Valiron kindly
calculated for us the structure of C$_5$N$^-$, and gave an estimate of its
dipole moment.  This was one of his last scientific contributions; he
died untimely on August 31$^{th}$.  Pierre was a valued colleague and
friend, who will be greatly missed.
We thank Carl Gottlieb for supporting laboratory frequencies.
We acknowledge funding support from Spanish MEC
trough grants AYA2006-14876 and ESP2004-665, from
PRICIT CM project
S-0505/ESP-0237 (ASTROCAM), and from FP6 MCTN
"The Molecular Universe". M. Ag\'undez also
acknowledges grant AP2003-4619 from Spanish MEC.
The work in Cambridge is supported by NSF grant
CHE-0701204 and NASA grant NNX08AE05G.

\begin{deluxetable}{lrrcc}
\tabletypesize{\scriptsize}
\tablecaption{Observed line parameters in IRC +10216\label{table1}}
\tablewidth{0pc}
\startdata \hline \hline
\multicolumn{1}{c}{}           & \multicolumn{1}{c}{Obs. Freq.} & \multicolumn{1}{c}{Cal. Freq.} & \multicolumn{1}{c}{$\int$$T_A^*$$dv$} & \multicolumn{1}{c}{v$_{\rm exp}$} \\
\multicolumn{1}{c}{Transition} & \multicolumn{1}{c}{(MHz)}      & \multicolumn{1}{c}{(MHz)}      & \multicolumn{1}{c}{(K km/s)}          & \multicolumn{1}{c}{(km/s)}        \\
\hline \multicolumn{5}{c}{B1389 (C$_5$N$^-$)} \\
\hline
J=29-28       & 80550.7(4)  & 80550.7        & 0.30(3)  & 14.6(8) \\
J=30-29       & 83328.0(3)  & 83328.0        & 0.23(3)  & 15.3(7) \\
J=32-31       & 88883.1(10) & 88882.7        & 0.20(3)  & 14.5$^a$\\
J=33-32       & 91660.0(6)  & 91660.0        & 0.25(3)  & 15.4(7) \\
J=34-33       & 94437.6(5)  & 94437.3        & 0.35(3)  & 15.6(8) \\
J=35-34       & 97214.7(3)  & 97214.5        & 0.21(5)  & 14.8(6) \\
J=36-35       & 99992.3(8)  & 99991.8        & 0.22(3)  & 15.1(8) \\
J=37-36       & 102769.5(10)& 102769.0       & 0.24(4)  & 14.5$^a$\\
J=38-37       & 105546.4(10)& 105546.1       & 0.17(3)  & 14.5$^a$\\
J=39-38       & 108323.0(5) & 108323.2       & 0.15(2)  & 14.5$^a$\\
J=40-39       & 111099.1(15)& 111100.4       & 0.16(3)  & 14.5$^a$\\
\hline \multicolumn{5}{c}{B1390} \\
\hline
J=59/2-57/2   & 82007.0(5)  & 82006.4        & 0.29(2)  & 16.8(9) \\
J=61/2-59/2   & 84786.4(2)  & 84786.5        & 0.23(2)  & 13.7(7) \\
J=65/2-63/2   & 90346.9(10) & 90346.9        & 0.14(2)  & 14.5$^a$\\
J=67/2-65/2   & 93127.0(10) & 93127.1        & 0.15(5)  & 14.5$^a$\\
J=69/2-67/2   & 95907.3(4)  & 95907.3        & 0.24(2)  & 14.5$^a$\\
J=73/2-71/2   & 101468.5(10)& 101467.9       & 0.23(4)  & 14.5$^a$\\
J=75/2-73/2   & 104248.2(10)& 104248.2       & 0.17(2)  & 14.5$^a$\\
J=77/2-75/2   & 107028.7(5) & 107028.6       & 0.29(2)  & 14.3(5) \\
J=79/2-77/2   & 109808.4(10)& 109809.0       & 0.09(3)  & 14.5$^a$\\
\hline \multicolumn{5}{c}{B1394} \\
\hline
N=29-28 a     & 80878.5(4)  & 80878.6        & 0.34(6)  & 14.2(5) \\
N=29-28 b     & 80889.5(4)  & 80889.9        & 0.34(6)  & 15.2(6) \\
N=30-29 a     & 83667.0(3)  & 83667.6        & 0.24(3)  & 14.5$^a$\\
N=30-29 b     & 83678.6(3)  & 83678.6        & 0.34(3)  & 14.5$^a$\\
N=31-30 a     & 86456.6(3)  & 86456.6        & 0.34(2)  & 14.4(3) \\
N=31-30 b     & 86467.5(3)  & 86467.4        & 0.42(2)  & 14.4(3) \\
N=32-31 a     & 89245.9(3)  & 89245.6        & 0.30(2)  & 14.8(2) \\
N=32-31 b     & 89255.9(3)  & 89256.1        & 0.33(2)  & 14.5$^a$\\
N=33-32 a     & 92034.1(10) & 92034.5        & 0.35(4)  & 14.5$^a$\\
N=33-32 b     & 92044.1(10) & 92044.7        & 0.41(4)  & 14.5$^a$\\
N=34-33 a     & 94823.4(10) & 94823.5        & 0.32(2)  & 14.5$^a$\\
N=34-33 b     & 94832.8(10) & 94833.4        & 0.33(2)  & 14.5$^a$\\
N=35-34 a     & 97612.8(20) & 97612.4        & 0.29(4)  & 14.5$^a$\\
N=35-34 b     & 97623.8(20) & 97622.0        & 0.26(4)  & 14.5$^a$\\
N=36-35 a     & 100401.8(5) & 100401.3       & 0.25(2)  & 14.5$^a$\\
N=36-35 b     & 100412.0(5) & 100410.5       & 0.27(2)  & 14.5$^a$\\
N=37-36 a     & 103189.6(5) & 103190.2       & 0.22(2)  & 14.5$^a$\\
N=37-36 b     & 103197.6(5) & 103199.1       & 0.30(2)  & 14.5$^a$\\
N=38-37 a     & 105978.6(8) & 105979.0       & 0.25(4)  & 14.2(5) \\
N=38-37 b     & 105986.1(8) & 105987.6       & 0.27(4)  & 14.9(5) \\
N=39-38 a     & 108767.3(15)& 108767.8       & 0.10(3)  & 14.5$^a$\\
N=40-39 a     & 111557.0(10)& 111556.6       & 0.14(3)  & 14.5$^a$\\
N=41-40 a     & 114344.9(6) & 114345.4       & 0.18(3)  & 14.5$^a$\\
N=41-40 b     & 114353.2(15)& 114352.9       & 0.17(5)  & 14.5$^a$\\
\hline \multicolumn{5}{c}{B1401} \\
\hline
J=59/2-57/2   & 82677.5(8)  & 82677.7        & 0.20(2)  & 14.5$^a$\\
J=61/2-59/2   & 85479.5(5)  & 85479.3        & 0.21(2)  & 14.5$^a$\\
J=63/2-61/2   & 88280.0(20) & 88280.8        & 0.24(4)  & 14.5$^a$\\
J=65/2-63/2   & 91082.0(2)  & 91082.2        & 0.33(3)  & 14.5$^a$\\
J=67/2-65/2   & 93883.1(10) & 93883.6        & 0.11(3)  & 14.5$^a$\\
J=69/2-67/2   & 96684.2(10) & 96684.7        & 0.09(2)  & 14.5$^a$\\
J=71/2-69/2   & 99486.0(10) & 99485.8        & 0.11(2)  & 14.5$^a$\\
J=73/2-71/2   & 102287.0(6) & 102286.8       & 0.19(3)  & 14.7(6) \\
J=75/2-73/2   & 105087.6(6) & 105087.6       & 0.10(2)  & 14.5$^a$\\
J=77/2-75/2   & 107889.0(10)& 107888.3       & 0.07(2)  & 14.5$^a$\\
\hline
\enddata
\tablecomments{
Number in parentheses are 1$\sigma$ uncertainties
in units of the last digits. A superscript ``a'' indicates that
the linewidth parameter v$_{\rm exp}$ has been fixed.}
\end{deluxetable}

\begin{deluxetable}{lcccc}
\tabletypesize{\scriptsize}
\tablecaption{Derived rotational constants\label{table-series}} \tablewidth{0pc}
\startdata \hline \hline
\multicolumn{1}{c}{Series} & \multicolumn{1}{c}{B(MHz)}
& \multicolumn{1}{c}{D(Hz)}     & \multicolumn{1}{c}{Nlines}& 
\multicolumn{1}{c}{J-range} \\
\hline
B1389  & 1388.860( 2) & 33(1)  & 13  &  8, 29-40\\
B1390  & 1389.878( 7) & -35(3) &  9  & 59/2-79/2\\
B1394  & 1394.609(10) & 32(4)  & 22  & 29-41$^*$ \\
B1401  & 1401.559(26) & 139(7) & 7 & 59/2-75/2
\enddata
\tablecomments{$^*$) B1394 corresponds to a $^2\Sigma$ vibronic
state with $\gamma$=15.2(13) MHz and $\gamma_d$=-1.53(35) KHz.
Number in parentheses are 1$\sigma$ uncertainties
in units of the last digits.}
\end{deluxetable}

\end{document}